\documentclass[notitlepage,letterpaper]{article}
\usepackage[ansinew]{inputenc} % Acepta caracteres en castellano
\usepackage[colorlinks=true,urlcolor=blue,linkcolor=blue,citecolor=blue]{hyperref} % navega por el doc
\usepackage{graphicx}
\usepackage{amsfonts}
\usepackage{amssymb}
\usepackage{color}
\usepackage{harvard}
\usepackage{cite}

\usepackage{amsthm}

\begin{document}

\title{Cracking isotropic and anisotropic relativistic spheres}
\author{
\textbf{Guillermo A. Gonz\'{a}lez}\thanks{\texttt{guillego@uis.edu.co}}, \textbf{Anamar\'ia Navarro}\thanks{\texttt{ana.navarro1@correo.uis.edu.co}} \\
\textit{ Escuela de F\'isica, Universidad Industrial de Santander,  }\\ 
\textit{A. A. 678, Bucaramanga 680002, Colombia}; \\
 and \textbf{Luis A. N\'{u}\~{n}ez}\thanks{\texttt{lnunez@uis.edu.co}} \\
\textit{ Escuela de F\'isica, Universidad Industrial de Santander,  }\\ 
\textit{Bucaramanga 680002, Colombia} and \\
\textit{Centro de F\'{\i}sica Fundamental, Departamento de F\'{\i}sica,} \\ 
\textit{Universidad de Los Andes, M\'{e}rida 5101, Venezuela.} 
}
\maketitle

\begin{abstract}
We explore the influence of density fluctuations on isotropic and anisotropic configurations, extending the concept of cracking for general relativistic fluid spheres. This concept, conceived to describe the behaviour of anisotropic matter distributions just  after its departure from equilibrium, could provide some insight on potential instabilities and future evolution of relativistic fluids. We have refined the idea of cracking, considering local fluctuations --represented by any function of compact support defined in a closed interval-- and their effect on the state variables and their gradients through  ``barotropic'' equations of state, $P = P(\rho)$  and $P_{\perp}= P_{\perp}(\rho)$. Under this approach it is found that both isotropic and anisotropic models could exhibit cracking (or overturning), and that previously crackable anisotropic models become uncrackable.
\end{abstract}

\section{Introduction}
The effects of gravitational instabilities on fluid bodies are one of the key problems in astrophysics and have been examined for centuries. Matter content of selfgravitating objects are macroscopically modeled by an equation of state, which relates their physical variables and attempts to describe the most important microscopic physical processes taking place within it. Thus, pathologies on the behavior of the equation of state under perturbations of its physical variables --like energy density and pressure-- are critical to understand the stability of the configuration.
  
Instabilities of general relativistic fluids spheres have been extensively considered under two complementary approaches: studying the evolution of dynamical perturbations and determining convective instabilities occurs within a matter distribution (see \cite{FriedmanStergioulas2014} and reference therein).  The first approach --starting from the seminal works of S. Chandrasekhar, R.F. Tooper and J.M. Bardeen \cite{Chandrasekhar1964a,Chandrasekhar1964b, Tooper1964b,Tooper1965,Bardeen1965} and formalized a decade later by  J. L. Friedman and B. F. Schutz \cite{FriedmanSchutz1975}-- concerns to the study of the evolution of perturbations of the physical variables through the matter distribution. The second scheme applies the buoyancy principle which leads that pressure and energy density most both, decrease outwards in any hydrostatic matter configuration\cite{Bondi1964B,Thorne1966,Kovetz1967}.  

An other approach to study the behavior of a fluid distributions just after it departs from its equilibrium, was proposed by L. Herrera and collaborators\cite{Herrera1992,DiPriscoEtal1994,DiPriscoHerreraVarela1997,HerreraFuenmayorLeon2016}. They study the tidal acceleration of contiguous fluids elements generated by a perturbation on the physical variables and showed that it is possible to identify a total radial force distribution that changes its sign within the system. They associate this change of sign to a pathology of the equation of state, that could present a cracking (or overturning) induced by local anisotropy (unequal stresses) of the fluid or, by the emission of radiation in the free streaming out limit, which also generates anisotropic pressure distribution. 

Herrera and coworkers assuming  simultaneous constant perturbations of the density and on the anisotropy of local pressures, studied ``snapshots'' of anisotropic systems just after they leave the equilibrium (or within a timescale shorter than the hydrostatic one) and identify cracking whenever a radial force directed inward, changes its sign pointing outward.  Mimoso and co-workers --studying the existence and stability of a dividing shells, separating expanding and collapsing regions in spherically symmetric spacetimes with anisotropic fluids-- have shown that the cracking concept is closely related to the problem of structure formation\cite{MimosoLeDelliouMena2010,MimosoLeDelliouMena2013,LeDelliouEtal2013}.  Other contribution to the general cracking  framework considers anisotropic fluids having ``barotropic'' equations of state for radial and tangential pressure profiles, $P = P(\rho)$  and $P_{\perp}= P_{\perp}(\rho)$, and explore the effects of constant density perturbations on the gravitational force distribution within anisotropic matter configuration  \cite{AbreuHernandezNunez2007b}. These  authors present cracking viability associated to the sign of the difference between the tangential and the radial sound speed, and this criterium has been applied to several astrophysical scenarios\cite{RahamanEtal2010,RahamanEtal2012,KalamEtal2012,KalamEtal2014,AzamMardanRehman2015A,AzamMardanRehman2015B}. 
 
The above cracking formalism has two important assumptions. First, as we have mentioned above, the system is perturbed by two (in density and in local anisotropy) constant, simultaneous and independent perturbations.
Second, that these perturbations do not change the gradient of pressure, i.e. that under perturbations of density and local anisotropy, the radial pressure of the system maintains the same radial dependence it had in equilibrium.   

In this work we extend the cracking idea, assuming a barotropic equation of state, $P=P(\rho)$, affected by non constant (local) density fluctuations --represented by any function of compact support defined in a closed interval-- affecting physical state variables and their gradients\cite{GonzalezNavarroNunez2015}. 

This paper is organized as follows: Section \ref{InstabCracking} will describe the concept of cracking for selfgravitating anisotropic matter configurations; in Section \ref{EffectsAnisotropic} we present the effects of local density fluctuations on the force distributions within isotropic/anisotropic matter configurations and we illustrate the effects with some examples and some conclusions are displayed in Section \ref{Remarks}.

\section{Cracking of anisotropic relativistic spheres}
\label{InstabCracking}
In a series of papers Herrera and coworkers  
\cite{Herrera1992,DiPriscoEtal1994,DiPriscoHerreraVarela1997} introduced the concepts of cracking and overturning to describe the behaviour of selfgravitating  anisotropic matter configurations -just after its departure from equilibrium- when the radial force reverses its sign beyond some value of the radial coordinate within the configuration. They considered an static spherically symmetric metric,  
\begin{equation}
\mathrm{d}s^2 = \mathrm{e}^{2 \nu(r)}\,\mathrm{d}t^2-\mathrm{e}^{2\lambda(r)}\,\mathrm{d}r^2-r^2 \left(\mathrm{d}\theta ^2+
\sin^2\theta\,\mathrm{d}\phi^2\right),
\label{metricSpherical}
\end{equation}
and an anisotropic fluid, 
\begin{equation}
{T}_{\mu \nu} = (\rho + P_{\perp}){{u}}_\mu{ {u}}_\nu - P_{\perp}{g}_{\mu \nu}  +
(P-P_{\perp}){{v}}_\mu {{v}}_\nu   \,,\label{tmunu}
\end{equation}
where ${{u}}_\mu =  (\mathrm{e}^{\nu}, 0, 0, 0)$, ${{v}}_\mu =  (0,-\mathrm{e}^{\lambda}, 0, 0)$, 
$\rho$ describes the energy density, $P$ the radial pressure, and $P_{\perp}$ the tangential pressure of the fluid. 

The hydrostatic equilibrium equation describing this anisotropic matter configuration is  
\begin{equation}
 \mathcal{R} = \frac{\mathrm{d} P}{\mathrm{d} r} +(\rho +P)\frac{m + 4 \pi r^{3}P}{r(r-2m)}-\frac{2(P_\perp -P)}{r} =0 ,
 \label{Ranitov}
\end{equation}
but it also represents the distribution of forces in the system and evaluated at the moment immediately after a perturbation may be different than zero. This implies that a distribution of forces may appear on the system driving cracking (or overturning) at some value of $0 < r_{\mathcal{C}} < R$ (where $R$ denotes the boundary of the distribution), when $\mathcal{R}$ changes its sign within the configuration. Cracking appears when the perturbed force distribution change its sign, $\delta  \mathcal{R} < 0  \rightarrow \delta \mathcal{R} > 0$, i.e. at the inner part of the sphere the force pointing inward, reverses its sign at a given point $r_{\mathcal{C}}$. On the other hand, overturning occurs when the net force, directed outward, inverts its sign as  $\delta \mathcal{R} > 0  \rightarrow \delta \mathcal{R} < 0$ at $r_{\mathcal{C}}$. 

\section{The effects of local density perturbations}
\label{EffectsAnisotropic}
In this section we show the consequences of density perturbations $\delta \rho$ on the stability of an anisotropic matter configuration with two barotropic equations of state, $P = P(\rho)$  and $P_{\perp}= P_{\perp}(\rho)$, which are the simplest type of equations of state, and their existence is useful to describe the properties of matter. We assume that density fluctuations induce variations into all other physical variables, i.e. $m(r), P(r)$ and $P_{\perp}(r)$, and their derivatives;  affect the hydrostatic equilibrium of the system and lead to a non-vanishing total radial force distribution ($\delta \mathcal{R} \neq 0$) within the  configuration. 

\subsection{Local density perturbations}
Note that a density perturbation, $\rho  \rightarrow \rho + \delta \rho,$ induces a perturbation of their gradient,
\begin{equation}
\rho' (\rho + \delta \rho) \approx \rho' (\rho) + \delta \rho' = \rho' (\rho) + \frac{\mathrm{d} \rho'}{\mathrm{d} \rho} \delta \rho,
\end{equation}
where primes stand for radial differentiation, which must be consistent with the expression
\begin{equation}
\frac{d }{dr} \left[ \rho + \delta \rho \right] = \rho' (\rho) + \delta \rho' = \rho' (r) + \frac{\mathrm{d \ } }{\mathrm{d} r} \delta \rho,
\end{equation}
in such a way that we can exchange the prime and the $\delta$ as
\begin{equation}
\delta \rho' = \frac{\mathrm{d } \rho' }{\mathrm{d} \rho} \delta \rho = \frac{\mathrm{d \ } }{\mathrm{d} r} \delta \rho \, .
\end{equation}
To have a consistent perturbation schema, we must to consider local perturbations of density, that can be properly described by any function of compact support, $\delta \rho = \delta \rho(r)$, defined in a closed interval $\Delta r \ll R$, with $R$ is the total radius of the configuration. 

Accordingly, these local density perturbations generate fluctuations in mass, radial pressure, tangential pressure and pressure gradient, that can be represented up to linear terms in density fluctuation as: 
\begin{eqnarray}
 P(\rho + \delta \rho)   & \approx   P(\rho) + \delta P  \approx & P(\rho) + \frac{\mathrm{d} P}{\mathrm{d} \rho} \delta \rho \, ,  \\ \nonumber \\ 
 P_{\perp}(\rho + \delta \rho)   & \approx   P_{\perp}(\rho) + \delta P_{\perp}  \approx & P_{\perp}(\rho) + \frac{\mathrm{d} P_{\perp}}{\mathrm{d} \rho} \delta \rho \, , \\ \nonumber \\ 
 P'(\rho + \delta \rho)   & \approx   P'(\rho) + \delta P'  \approx & P'(\rho) + \frac{\mathrm{d} P'}{\mathrm{d} \rho} \delta \rho \; \;\mathrm{and} \\ \nonumber \\ 
 m(\rho + \delta \rho)   & \approx   m(\rho) + \delta m  \approx & m(\rho) + \frac{\mathrm{d} m}{\mathrm{d} \rho} \delta \rho  \, ;
\end{eqnarray}
where the last part in each equation represents the perturbed variables calculated by Taylor expansion. Thus, we can clearly identify the perturbed variables as the corresponding part in the linear Taylor series as: 
\begin{eqnarray} \label{deltas}
 \delta P &=& \frac{\mathrm{d}P}{\mathrm{d} \rho} \delta \rho =  v^2 \delta \rho \, , \\   
\nonumber \\
 \delta P_\perp &=& \frac{\mathrm{d} P_{\perp}}{\mathrm{d} \rho} \delta \rho =  v_\perp^2 \delta \rho \, ,  \\
\nonumber \\
\delta P' &=& \frac{\mathrm{d} P'}{\mathrm{d} \rho} \delta \rho  =   \frac{\mathrm{d \ }}{\mathrm{d} \rho} \left[ \frac{\mathrm{d} P}{\mathrm{d}r} \right] \delta \rho =   \frac{\mathrm{d}}{\mathrm{d}\rho} \left[ \frac{\mathrm{d}P}{\mathrm{d}\rho} \frac{\mathrm{d}\rho}{\mathrm{d} r} \right]\delta \rho  \; , \nonumber  \\  \nonumber \\
&=&  \frac{\mathrm{d}}{\mathrm{d}\rho} \left[v^2 \rho' \right]\delta \rho = \frac{1}{ \rho'} \frac{\mathrm{d}}{\mathrm{d}r} \left[v^2 \rho' \right]\delta \rho =  \left[  (v^2)' + v^2\frac{\rho''}{\rho'} \right] \delta \rho =   \frac{P''}{\rho'}  \delta \rho \;  \; \; \mathrm{and} \\  \nonumber \\
\delta m &=& \frac{\mathrm{d} m}{\mathrm{d} \rho} \delta \rho =  \frac{\mathrm{d} m}{\mathrm{d} r} \left( \frac{\mathrm{d} r}{\mathrm{d} \rho} \right) \delta \rho   =  \frac{m'}{ \rho'} \delta \rho = \frac{4 \pi r^2 \rho}{\rho'}\delta \rho   \label{deltam} \; ;
\end{eqnarray}
where 
\begin{equation}
 v^{2} = \frac{\mathrm{d} P }{ \mathrm{d} \rho}, \quad 
 v^{2}_{\perp} = \frac{\mathrm{d} P_{\perp}  }{ \mathrm{d} \rho}, \quad
\mathrm{and} \quad
 m = 4\pi \int_0^{r} \rho(\bar{r})\bar{r}^2 \mathrm{d}\bar{r} \, ,  \label{def_masa}
\end{equation}
denote radial sound speed, tangential sound speed, and mass, respectively. 

Notice that our perturbation scenario differs from the original one presented by Herrera and collaborators 
\cite{Herrera1992,DiPriscoEtal1994,DiPriscoHerreraVarela1997}, where fluctuations in density and anisotropy were considered  independent and simultaneous, and it is also different from the other previous study \cite{AbreuHernandezNunez2007b} because our approach assumes non constant density fluctuations which affects the pressure gradient, i.e. $\rho + \delta \rho(r) \rightarrow  P^{\prime}(\rho) + \delta P^{\prime} $.

Following the above guidelines, we formally expand equation (\ref{Ranitov}) as
\begin{equation}
\label{RAniExpanded}
\mathcal{R} \approx \mathcal{R}_{0}(\rho, P, P_{\perp}, P^{\prime}, m) + {\delta \mathcal{R}},
\end{equation}
and comparing it with the corresponding Taylor expansion, we have
\begin{equation}
{\delta \mathcal{R}} =
\frac{\partial \mathcal{R}}{ \partial \rho} \delta \rho
+\frac{\partial \mathcal{R}}{ \partial P} \delta P
+\frac{\partial \mathcal{R}}{ \partial P_{\perp}} \delta P_{\perp}
+\frac{\partial \mathcal{R}}{ \partial P^{\prime}} \delta P^{\prime}
+\frac{\partial \mathcal{R}}{ \partial m} \delta m ,
\end{equation}
where
\begin{equation}
\mathcal{R}_{0}(\rho, P, P_{\perp}, P^{\prime}, m) = 0
\end{equation}
because initially the configuration is in equilibrium. Thus, we can write
\begin{equation}
\label{deltaRAni}
 \delta \mathcal{R} = \delta \rho \left\{ 
 \frac{\partial \mathcal{R} }{\partial \rho}  
 + \frac{\partial \mathcal{R}}{\partial P}v^2  
 + \frac{\partial \mathcal{R}}{\partial P_\perp}  v_\perp^2 
 + \frac{\partial \mathcal{R}}{\partial m}  \frac{4 \pi r^2 \rho}{\rho'}    
 + \frac{\partial \mathcal{R}}{\partial P'}  \frac{P''}{\rho'} 
\right\}.  
\end{equation}
As can be easily calculated from (\ref{Ranitov}), the derivatives of the force distribution, $\mathcal{R}$, are given by
\begin{eqnarray}
\label{partialRAni}
\frac{\partial \mathcal{R}}{\partial \rho} & = & \frac{  4 \pi r^3 P + m  }{ r(r - 2m)} ,  \\ \nonumber \\  
\frac{\partial \mathcal{R}}{\partial m} & = & \frac{ (\rho + P)( 1 + 8\pi r^2 P)  }{(  2 m  - r )^2 } ,  \\ \nonumber \\
\frac{\partial \mathcal{R}}{\partial P} & = & \frac{ m + 4 \pi r^3 ( \rho + 2 P )}{r(r-2m)}  + \frac{2}{r} , \\ \nonumber \\
\frac{\partial \mathcal{R} }{\partial P_\perp} & = & -\frac{2}{r} \qquad \mathrm{and} \qquad \frac{\partial \mathcal{R}}{\partial P'} = 1 \; .
\end{eqnarray}

To identify the possible sources of sign change, we can write (\ref{deltaRAni}) as
\begin{equation}\label{deltaRAni2}
\delta \mathcal{R} = \delta \rho \left\{ {\widetilde{\mathcal R}_{1}} + {\widetilde{\mathcal R}_{2}} + {\widetilde{\mathcal R}_{3}} + {\widetilde{\mathcal R}_{4}} + {\widetilde{\mathcal R}_5} + {\widetilde{\mathcal R}_{6}} \right\},
\end{equation}
where
\begin{eqnarray}
{\widetilde{\mathcal R}_{1}} & = & \frac{m + 4 \pi  r^3 P}{r(r - 2m)},  \\ \nonumber \\
{\widetilde{\mathcal R}_{2}} & = & \left[ \frac{(\rho + P) ( 1 + 8 \pi r^2 P)}{(r - 2m)^2}  \right]  \frac{4 \pi}{3}r^{3},  \\ \nonumber \\
{\widetilde{\mathcal R}_{3}} & = & \left[ \frac{ m + 4 \pi r^3 ( \rho + 2 P )}{r(r-2m)} \right] v^{2},    \\ \nonumber \\
{\widetilde{\mathcal R}_{4}} & = & 2 \left[ \frac{v^{2} -v^{2}_{\perp} }{r} \right],  \qquad
{\widetilde{\mathcal R}_{5}}   = \frac{P''}{\rho'} , \quad \mathrm{and} \\ \nonumber \\
{\widetilde{\mathcal R}_{6}} & = & \left[ \frac{ (\rho + P)( 1 + 8\pi r^2 P  ) }{(  2 m  - r )^2 } \right] \left( \frac{\rho}{\rho'} - \frac{r}{3}\right) 4 \pi r^2 ,
\end{eqnarray}

Observe this picture describes the behaviour of the fluid just  after its departure from equilibrium and that the expression for the perturbation of the net force (\ref{deltaRAni2}) is independent of the arbitrary functional form of any density perturbation represented by a function of compact support defined in an interval $\Delta r \ll R$ and notice that: 
\begin{enumerate}
\item any perturbation at the center of the matter distribution (or at the surface $r = R$) should vanish in order that the spherical symmetry be preserved,
\begin{equation}
\left. \delta \rho \right|_{r=0} \equiv 0
\quad \Rightarrow \quad
\left.\delta R\right|_{r=0} =0 \; ; 
\end{equation}  

\item any change of sign for $\delta \mathcal{R} $ should emerge from the last three terms because the first three are always be positive. The last term, ${\widetilde{\mathcal R}_{6}}$, is always negative if the buoyancy principle, i.e. $\rho'<0$ and $P'<0$, is invoked. Thus, ${\widetilde{\mathcal R}_{4}}$ and ${\widetilde{\mathcal R}_{5}}$ can change their signs, depending on the particular equation of state considered;

\item if these last three terms are negative but not big enough to counterbalance the first three, the configuration does not exhibit any cracking and will be stable under this type of density perturbations;

\item the contribution of the first four terms were considered in \cite{AbreuHernandezNunez2007b} for constant density perturbations. In the present work the last two terms emerge as a consequence of a variable density perturbation affecting both the pressure gradient and the mass function, and this is evident;

\item  equation (\ref{deltaRAni2}) admits cracking (or overturning) instabilities for isotropic matter distributions, when  $P(r)=P_{\perp}(r)$, and can be written as
\begin{equation}
\label{deltaRIso}
\delta \mathcal{R}_{\mathrm{iso}} = \delta \rho \left\{ {\widetilde{\mathcal R}_{1}} + {\widetilde{\mathcal R}_{2}} + {\widetilde{\mathcal R}_{3}} + {\widetilde{\mathcal R}_5} + {\widetilde{\mathcal R}_{6}} \right\},
\end{equation}
as the term ${\widetilde{\mathcal R}_{4}}$ vanishes, and it will be study in the next section.  
 
\end{enumerate}

\subsection{Density perturbation on isotropic spheres}
%%%%%% begin New
As we have mentioned above, the most important consequence of the present approach is that isotropic matter configurations exhibit cracking and/or overturning. To illustrate it, we present some examples of the effects of density perturbations on the cracking of isotropic matter configurations: first we consider an exact isotropic model \cite{Mehra1966} and next we workout one of the standard forms for the relativistic polytropic equation of state having, as its newtonian limit, the usual relation   
\begin{equation}
\label{PolytropeNewt}
P = K \rho_{0}^{\gamma} \equiv K \rho_{0}^{1 +1/n}  
\end{equation}
where, $P$ is the isotropic pressure,  $\rho_{0}$ the (baryonic) mass density, $\gamma$ the polytropic exponent (or polytropic constant) and $n$ the polytropic index \cite{Tooper1964,Bludman1973,HerreraBarreto2013}.

%%%% end New
\subsubsection{Mehra Model}
We start examining a model proposed by Mehra \cite{Mehra1966}, considered physically viable by Delgaty and Lake\cite{DelgatyLake1998} and defined by density $\rho$ and pressure $P$ distributions as:
\begin{eqnarray}
\rho & = & \frac{ \rho_{c}(R^2 - r^2)(\sqrt{\alpha_1} + 2\sqrt{\alpha_2} )^2 }{R^2\left[\alpha_1 + 4\sqrt{\alpha_1} \alpha_2 + 40\pi R^2 \rho_{c} - 40 \pi R^2 \rho_{c} \cos{\left(\frac{z-z_1}{2}\right)}^2 \right]} \quad \mathrm{and} \\  \nonumber \\
P &= & \frac{1}{3} \frac{\rho_{c}\left( R^2 - r^2\right)}{R^2}, \label{eq_Modelo_Mehra_Pr}
\end{eqnarray}
where
\begin{eqnarray}
\alpha_1 & = & 225 - 240 \pi R^2 \rho_{c} \cos{\left(\frac{z-z_1}{2}\right)} \, , \\ \nonumber \\
\alpha_2 & = & R\sqrt{10\pi \rho_{c}}\sin{\left(\frac{z-z_1}{2}\right)}  \, , 
\\ \nonumber \\
z & = & \ln{\left( \frac{r^2}{R^2} - \frac{5}{6} + \sqrt{\frac{r^4}{R^4} - \frac{5r^2}{3R^2} + \frac{5}{8\pi R^2 \rho_{c}} } \right)} ,
\end{eqnarray}
and
\begin{equation}
z_1 = \ln{\left( \frac{1}{6} + \sqrt{\frac{5}{8\pi R^2 \rho_{c}} - \frac{2}{3}} \right)}, 
\end{equation} 
with $R^2< 9/10\pi \rho_{c}$. 

If $M = m(R)$ the central density, $\rho_{c}$, can be written as 
\begin{equation}
\rho_{c}  = \frac{15 M}{8\pi R^3} = \frac{15\mu}{16\pi R^2},
\end{equation}
with $\mu = {2M}/{R}$.

\begin{figure}[h]
\begin{center}
\includegraphics[width = 0.70 \textwidth]{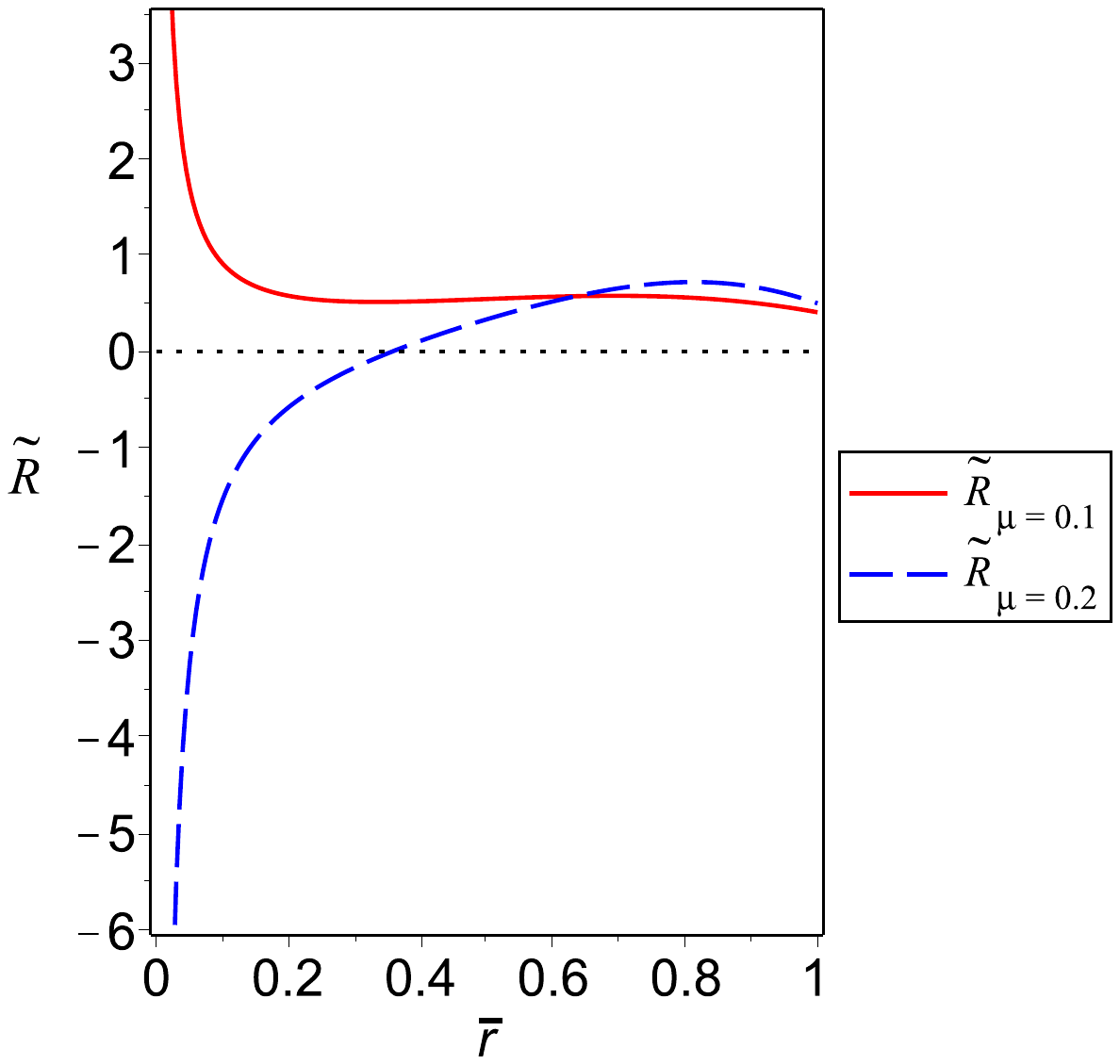}
\end{center}
\caption{$\widetilde{\mathcal{R}} = \delta \mathcal{R}/\delta \rho$ as a function of $\bar{r}= r/R$ for the isotropic Mehra-Model \cite{Mehra1966}  with $\mu = 0.1$ (solid line) and $\mu = 0.2$ (dashed line). The $\mu = 0.2$ curve presents a cracking point  $r_{\mathcal{C}}\approx 0.35$ which illustrates that cracking instability can be found for isotropic matter configurations.}\label{Grafica_Modelo_Mehra}
\end{figure}

Figure \ref{Grafica_Modelo_Mehra} displays the force distribution, ${\widetilde{\mathcal R}} = \delta \mathcal{R}/\delta \rho$, for the Mehra-model \cite{Mehra1966}  with two different values of the mass-radius $\mu$. Observe that the curve for $\mu=0.2$  changes its sign around $r\approx 0.35$, which illustrates that a cracking instability can be found for this isotropic matter configurations. 
%%%%%Figure

%%%% starting New
\subsubsection{Isotropic Polytrope}
\label{PolytropicExample}
Next, we workout the case of standard isotropic polytropes models which have been widely study in the literature since the 1960's \cite{Tooper1964,Bludman1973,NilssonUggla2000,MaedaEtal2002,HerreraBarreto2004,LaiXu2009,ThirukkaneshRagel2012} and recently extended to considere local anisotropic pressure distribution describing both newtonian and relativistic matter distributions \cite{HerreraBarreto2013B,HerreraBarreto2013}. 

To explore the effects of density perturbations on the cracking of isotropic polytropic matter configurations, we shall examine the polytropic equation of state of the form 
\begin{equation}
\label{PolitropCase2}
P = K \rho^{\gamma} \equiv K \rho^{1 +1/n} \, .
\end{equation}  
This one of the two relativistic polytropic equations of state which has, as its newtonian limit equation (\ref{PolytropeNewt})  (see the Case 2 in references \cite{HerreraFuenmayorLeon2016,HerreraBarreto2013}). 

The family of static isotropic politropes are described by the solutions of the usual system of stelar structure equations:
\begin{equation}
 \frac{\mathrm{d} P(r)}{\mathrm{d} r} = -(\rho(r) +P(r))\frac{m(r) + 4 \pi r^{3}P(r)}{r(r-2m(r))}  \quad \mathrm{and} \quad
  \frac{\mathrm{d} m(r)}{\mathrm{d} r} = 4 \pi r^{2}\rho(r) \, .
 \label{StructureEquations}
\end{equation}
Now changing variables as
\begin{equation}
\label{NewVariables}
\psi(\xi)^{n} = \frac{\rho}{\rho_{c}} , \;
v(\xi) = \frac{m(r) A^{3}}{4 \pi \rho_{c}} , \; 
\xi = r A , \; \;
A^{2} =  \frac{4 \pi \rho_{c}}{\alpha(n+1)} \;,   \; \; \alpha = P_{c}/ \rho_{c} \, ,
\end{equation}
and  substituting (\ref{PolitropCase2}) into the system (\ref{StructureEquations}) a new system is obtained:
\begin{equation}
\label{StructureEquationsPolyt}
\frac{ \mathrm{d}\psi( \xi) } {{\rm d} \xi} = {\frac { \left( 1+ \alpha \psi(\xi) \right)  \left( v  +\alpha \xi^{3} \psi( \xi)^{n+1} \right) }{\xi\, \left( 2 (n+1) \alpha v \left( \xi \right) -\xi \right) }}
\quad \mathrm{and} \quad 
\frac{ \mathrm{d}v } {{\rm d} \xi} = \xi^{2} \psi( \xi)^{n} \, ,
\end{equation}
with equation (\ref{PolitropCase2}) translated into
\begin{equation}
\label{PolitropePsi}
\frac{P}{\rho} = K \rho^{1/n} \equiv K \left( \rho_{c} \psi^{n} \right)^{1/n} = K_{1}\psi \, ,
\end{equation}
which lead to the generalized Lane-Emden equation, parametrized by the polytropic index, $n$, the pressure-density at the center, $\alpha$ and $K_{1}$ a dimensionless parameter relating pressure and density from the equation of state.    \cite{Tooper1964,Bludman1973,HerreraBarreto2013}.

To evaluate the existence of cracking or overturning within the isotropic politrope we substitute variables (\ref{PolitropCase2}), (\ref{NewVariables}) and (\ref{PolitropePsi}) into equation (\ref{deltaRIso}) to obtain
\begin{equation}
\label{Rtilede1}
\frac{\tilde{R}_{1}}{A} ={\frac { (n+1) \alpha}{\xi\, \left(\xi -2v \alpha n -2 v \alpha \right) } 
\left( K_{1}  \psi^{n+1}  \xi^{3}  +v  \right) }
\end{equation}
%%%%%
\begin{equation}
\label{Rtilede2}
\frac{\tilde{R}_{2}}{A} = \frac {\alpha  (n+1)  \xi^{3} \psi^{n}(K_{1}\psi +1) }{3 \left(\xi -2v \alpha n -2 v \alpha \right)^{2}}
\left( 2K_{1} \psi^{n+1}\alpha \xi^{2}(n+1) +1 \right) 
\end{equation}
%%%%%
\begin{equation}
\label{Rtilede3}
\frac{\tilde{R}_{3}}{A} = {\frac { (n+1) ^{2}K_{1} \psi \alpha}{n\xi\, \left(\xi -2v \alpha n -2 v \alpha \right)  } 
\left( 2K_{1}  \psi^{n+1}  \xi^{3}  +\psi^{n}  \xi^{3} +v  \right) }
\end{equation}
\begin{equation}
\label{Rtilede5}
\frac{\tilde{R}_{5}}{A} ={\frac{ (n+1) K_{1}  \left(  \left( \psi^{\prime}   \right) ^{2}n+\psi \psi^{\prime \prime}   \right) }{\psi^{\prime}  n}}
\end{equation}
\begin{equation}
\label{Rtilede6}
\frac{\tilde{R}_{6}}{A}  = \frac { \xi^{2}\alpha  (n+1)  \psi^{n}(K_{1}\psi +1) (\psi^{\prime}\xi n -3\psi)}
			{3 \psi^{\prime}  n  \left(\xi -2v \alpha n -2 v \alpha \right)^{2}} 
			\left( 2K_{1} \psi^{n+1}\alpha \xi^{2}(n+1) +1 \right) 
\end{equation}
where 
\begin{equation}
\label{PsiDerivatives}
\psi^{\prime} = \frac{ \mathrm{d}\psi( \xi) } {{\rm d} \xi}  \quad \mathrm{and} \quad \psi^{\prime \prime} = \frac{ \mathrm{d}^{2}\psi( \xi) } {{\rm d} \xi^{2}},  
\end{equation}
corresponding to equation (\ref{StructureEquationsPolyt}) and its derivative, respectively.

We integrate numerically the above system (\ref{StructureEquationsPolyt}) with the obvious initial conditions ($\psi(0)=1$ and $v(0)=0$), for wide variety of parameters ($n, \alpha, K_{1}$) and  stopping the integration when  the density vanishes and defining, at this point the total mass $v_{tot}$ and the corresponding external radius of the configuration $\xi_{tot}$). 

\begin{figure}
\begin{center}
\includegraphics[width=4in]{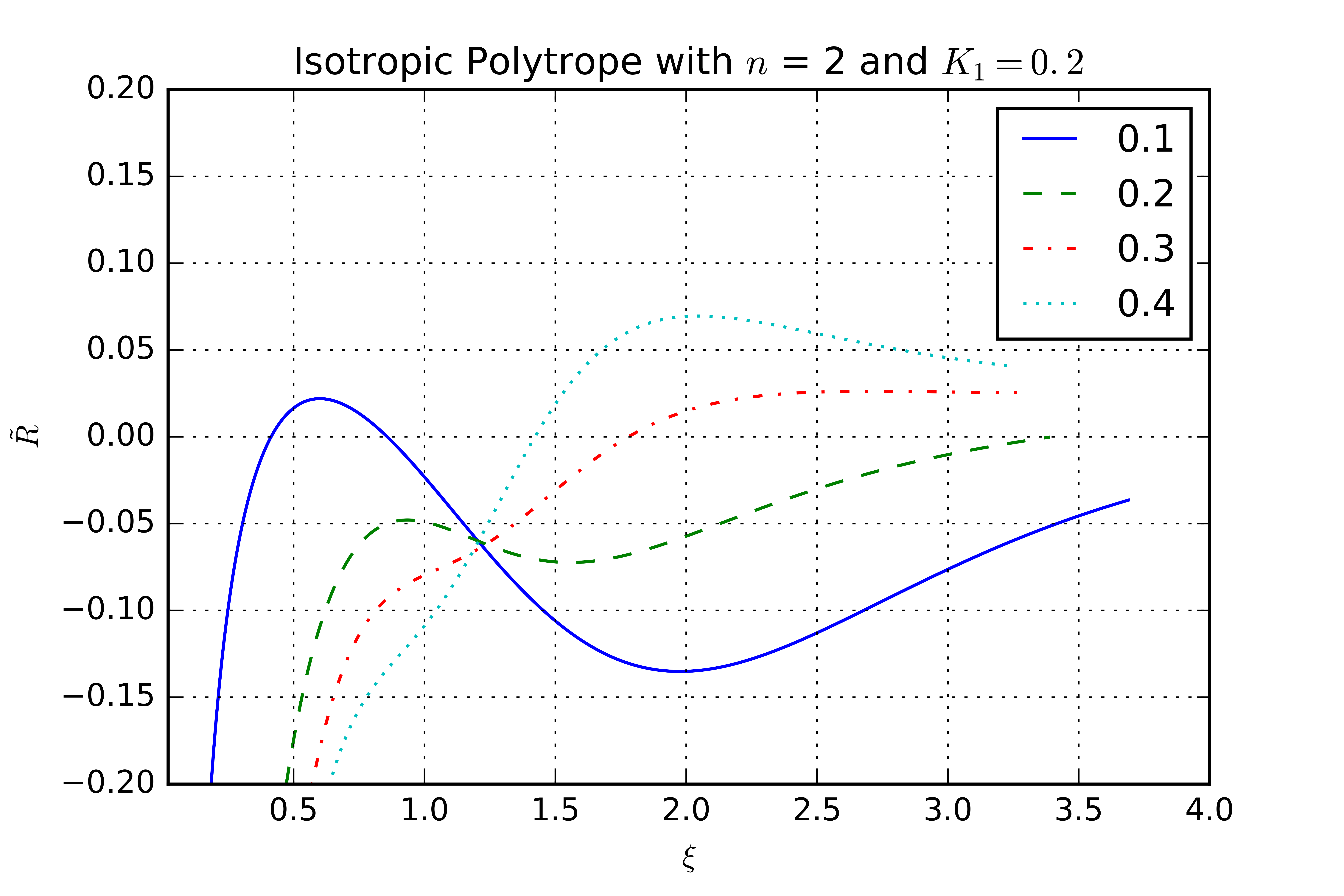}
\caption{$\widetilde{\mathcal{R}} = \delta \mathcal{R}/\delta \rho$, as a function of $\xi$, for the isotropic polytrope, with $n=2$, $K_{1} = 0.2$, and four different values of $\alpha = P_{c}/\rho_{c} = 0.1$ displayed by a  solid line; $\alpha = 0.2$ with ; $\alpha = 0.3$ dotted-dashed and $\alpha = 0.4$ dotted line. Notice the three different cases: cracking for $\alpha = 0.4$ and $\alpha = 0.3$; no cracking/no overturning $\alpha = 0.2$ and, finally cracking and overturning with $\alpha = 0.2$ (dashed). Observe that the four curves end at different $\xi_{tot}$.}
\label{CrackingIsoPolyt}
\end{center}
\end{figure}

Figure \ref{CrackingIsoPolyt} illustrates this effect through a particular set curves - for $n = 2$ and $K_{1}=2$- which represents three different cases for distinct values of $\alpha = P_{c}/\rho_{c}$: cracking for $\alpha = 0.4$ and $\alpha = 0.3$; no cracking/no overturning with $\alpha = 0.2$ and finally, cracking and after overturning for $\alpha = 0.1$. 

\subsection{Density perturbation on anisotropic spheres}
\begin{figure}[h]
\begin{center}
\includegraphics[width = 0.70 \textwidth]{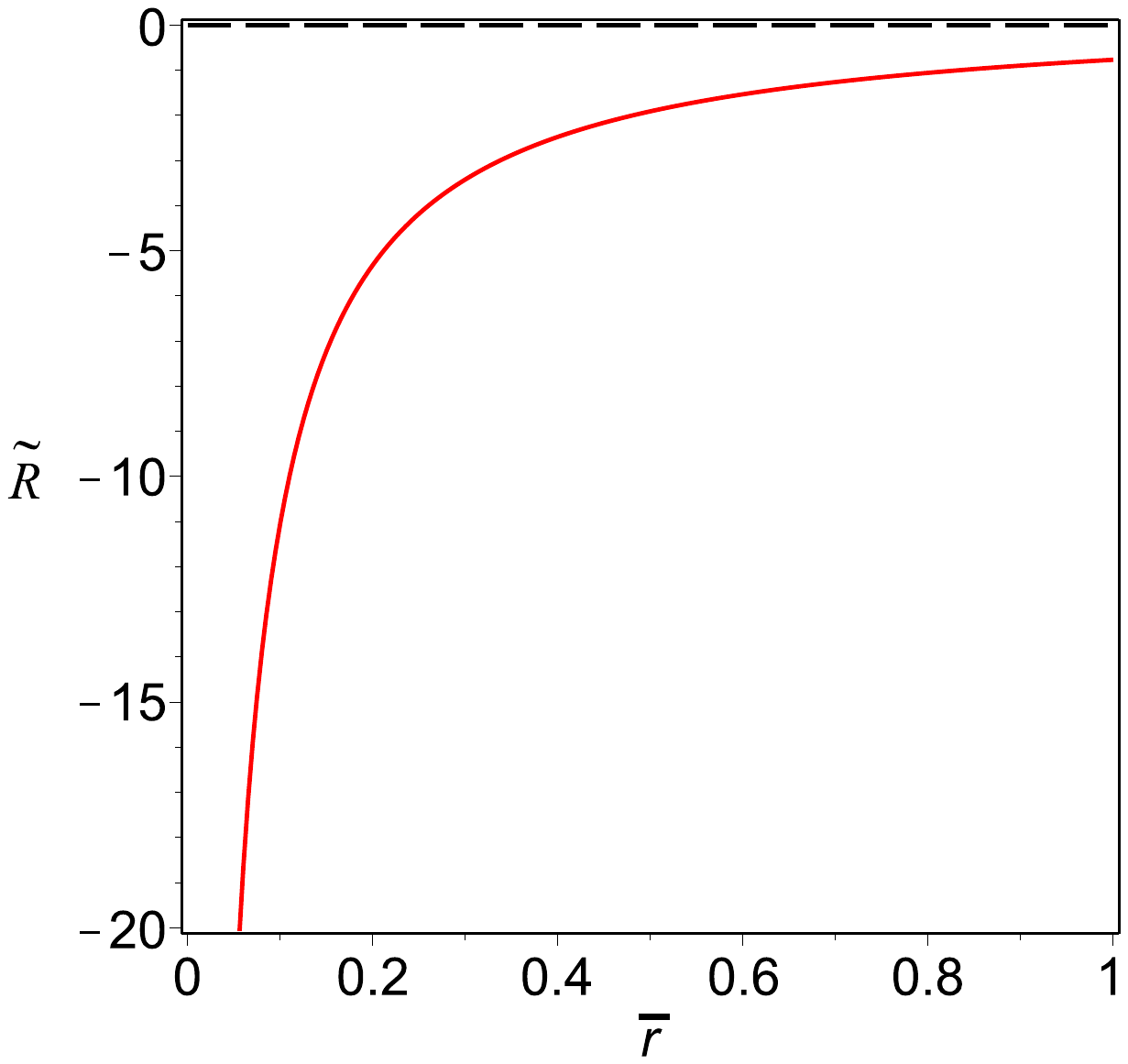}
\caption{$ \widetilde{\mathcal{R}} = \delta \mathcal{R}/\delta \rho$, for the anisotropic Herrera-Model \cite{Herrera1992}. $\widetilde{\mathcal{R}}$ does not change its sign and does not exhibit cracking instabilities. This picture differs from the one presented in \cite{Herrera1992}.} 
\label{Caradelamoneda_dR.pdf}
\end{center}
\end{figure}

In this section we present an analysis of two anisotropic models that were considered unstable in previous works. First, we will examine a model considered in \cite{Herrera1992}, described by
\begin{equation}
\label{eq_Model_Caradelamoneda} 
\rho =  \frac{K}{r^2}, \quad P = \frac{K}{3r^2 } \left( \frac{7 -  9 r^{-1/2}  }{1 - 3r^{-1/2}} \right) \quad \mathrm{and} \quad P_\perp =   P - \frac{\gamma}{r^2} 
\end{equation}
which proven to be crackable when simultaneous density and anisotropic perturbations take place. As it can be appreciated from Figure \ref{Caradelamoneda_dR.pdf} the total distribution force $\widetilde{\mathcal{R}}$ does not change its sign, thus the model could be considered as potentially stable under the present criterion (and also is under the previous sound velocity schema presented in  \cite{AbreuHernandezNunez2007b}). 
%%%% Figure

\begin{figure}[h]
\begin{center}
\includegraphics[width = 0.70\textwidth]{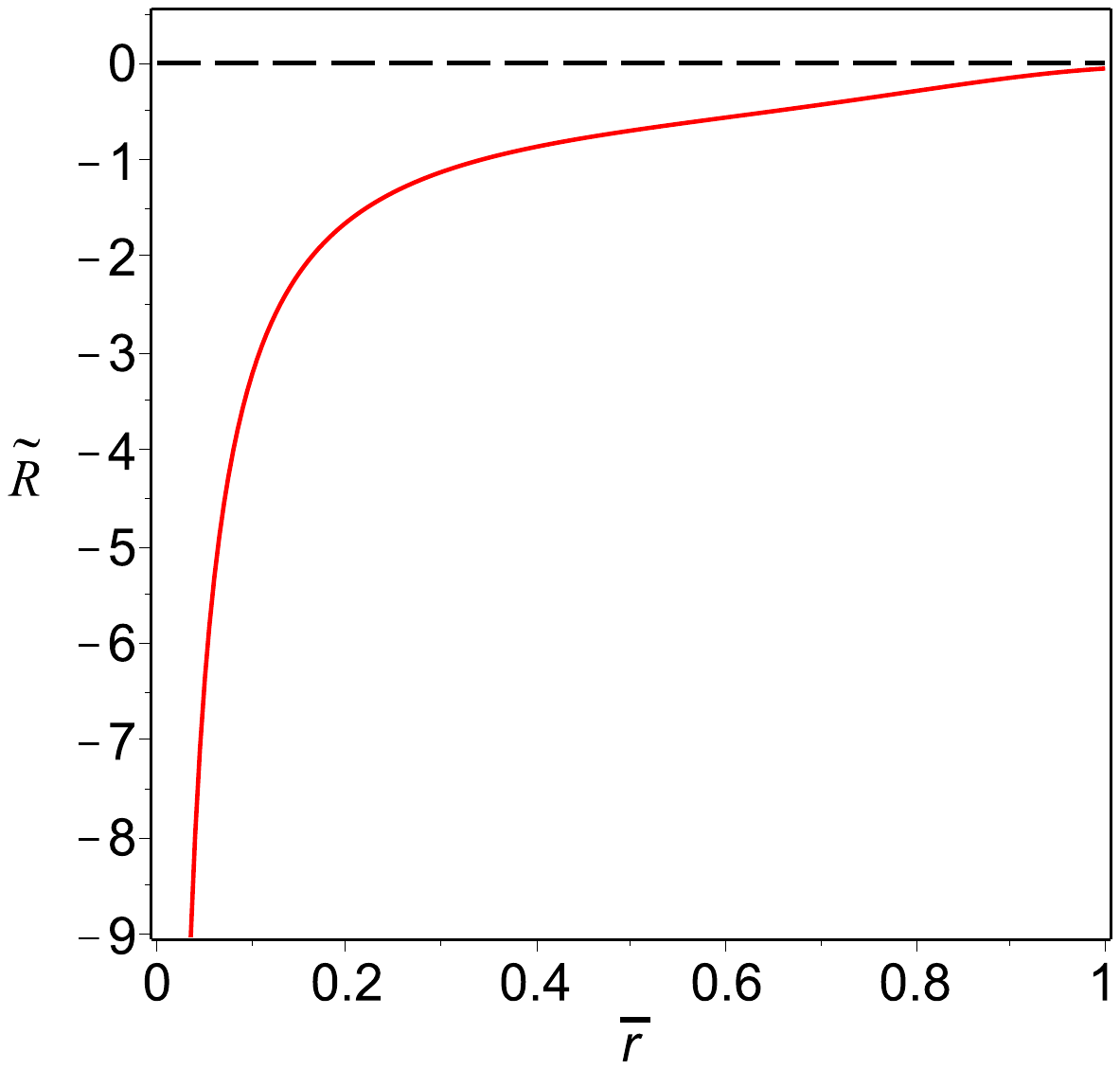}
\caption{$\tilde{\mathcal{R}} = \delta \mathcal{R}/\delta \rho$, for the anisotropic  Gokhroo/Mehra-Model  \cite{GokhrooMehra1994} with: $\mu = 0.42$, $K =3/56\pi$  and $\gamma = K/4$.  Observe that it does not present any cracking instability reported in \cite{AbreuHernandezNunez2007b}. }\label{Grafica_Modelo_Gokhromehra}
\end{center}
\end{figure}

The second model we study is based on the solution derived by Gokhroo and Mehra \cite{GokhrooMehra1994}, originally found by Florides\cite{Florides1974} and later by Stewart \cite{Stewart1982}, which is  described by
\begin{eqnarray}
\rho = \rho_{c} \left( 1 - \frac{Kr^2}{R^2} \right), \quad 
P =   \gamma \rho_{c} \left( 1 - \frac{2m}{r} \right) \left( 1 - \frac{r^2}{R^2}\right),  \quad \mathrm{and} \label{eq_Model_Gokhromera_rhoyPr} \\  \nonumber \\
P_\perp = P + \frac{1}{2}r P' + \frac{(\rho + P)(m + 4\pi r^3 P)}{2(r-2m)}, \label{eq_Model_Gokhromera_Pt} 
\end{eqnarray}
with the central density $\rho_{c}$ written as
\begin{equation}
\rho_{c} = \frac{15}{4\pi} \frac{M}{R^3(5-3K)} =\frac{15}{8\pi} \frac{\mu}{R^2(5-3K)}.
\end{equation}
This model studied in \cite{AbreuHernandezNunez2007b} displays cracking but, as it is clear from figure \ref{Grafica_Modelo_Gokhromehra}, with the present refinement of non-constant density perturbation - and assuming the same set of parameters: $\mu = 0.42$, $K =3/56\pi$  and $\gamma = K/4$- it does not present any cracking instability.
%%%% Figure

\section{Remarks and conclusions}
\label{Remarks}
We have extended the cracking formalism to consider local perturbations that affects pressure gradient. It is shown  that not only anisotropic matter distribution but also isotropic  configurations can experiment cracking when non constant (local) density fluctuations are considered.  As we have mentioned, the perturbations we assume are local, $\delta \rho = \delta \rho(r)$ --represented by any function of compact support defined in a closed interval $\Delta r \ll R$- and affect all physical variables and their gradients.

 The idea of cracking was originally conceived by L. Herrera\cite{Herrera1992} using independent and simultaneous constant perturbations, that may drive anisotropic matter configurations to exhibit cracking (or overturning).  Later, Abreu, Hern\'andez and N\'{u}\~{n}ez \cite{AbreuHernandezNunez2007b}, assuming fluids with  ``barotropic'' equations of state for radial and tangential pressures, $P = P(\rho)$  and $P_{\perp}= P_{\perp}(\rho)$,  show that constant (global) density perturbations could generate cracking on anisotropic relativistic fluids and this can be related to the sign of the difference between the tangential and radial sound speeds.  In their work, only  density fluctuations generate the perturbed scenario, affecting mass,  radial and  tangential pressure, but leaving unperturbed the pressure gradient; again in this density-driven-perturbation framework, only anisotropic distributions can exhibit cracking or overturning. 
 
We have extended, this density-driven-perturbation approach, assuming non-constant local density perturbations, which affect the gradient of pressure.  In our scheme, perturbations occur on a confined neighbourhood nearby a particular point within the distribution and therefore consider that they are more realistic. Additionally, the perturbation of the pressure gradient is also a suitable assumption because this gradient represents the distribution of gravitational forces within an hydrostatic configuration and should change with density fluctuations. Both extensions have proven to provide interesting outcomes because, we have obtained that isotropic matter distributions can also present cracking (or overturning) instability and it was clearly illustrated by two examples. It is particularly interesting the isotropic polytropic case shown in section \ref{PolytropicExample}, because it presents all the possible cracking scenarios for one model ($n=2$, $K_{1} = 0.2$) with distinct values of  $\alpha = P_{c}/\rho_{c}$: cracking for $\alpha = 0.4$ and $\alpha = 0.3$; no cracking/no overturning with $\alpha = 0.2$ and finally, cracking and overturning for $\alpha = 0.1$. Despite this polytropic model is only an example of the method, it should be pointed out that a similar cracking/overturning scenarios were also recently obtained by Herrera and collaborators for isotropic polytropic configurations (Figure 4 in reference \cite{HerreraFuenmayorLeon2016}), when density and isotropy were perturbed simultaneously.   

Clearly these two approaches generate distinct perturbation scenarios leading to different results and the above mentioned anisotropic Tolman VI model (\ref{eq_Model_Caradelamoneda}) is a clear example: it displays cracking when the independent and simultaneous perturbation of density and anisotropy is considered, but it does not shown any, for the density driven approach.  It is not clear which of the two scenarios is more likely to occur: the  simultaneous two-perturbation original scenario of Herrera or the density-driven framework, but surely both generate instabilities on relativistic matter configurations that should be evaluated.  

It is worth to be mentioned that, the idea of cracking -introduced to describe the behaviour of fluid distributions just  after their departure from equilibrium-  identifies tendencies of the fluid to split (or to compress) in time scales of the order of the hydrostatic time, $\tau _{hyd}\sim  \sqrt{r^3/m}$, which for neutron stars is of the order of $10^{-4}s$. In this ``snapshot'' of the system out of the equilibrium, it is shown that two nearby fluid elements can be accelerated, with respect to each other, at both sides of the cracking/overturning point. This effect  was initially only associated with anisotropic fluids, but now we have shown that it is possible to obtain this perturbed scenarios also for isotropic matter configurations.

It is necessary to study how cracking is related to the standard perturbation schema and under which circumstances these ``critical'' points can evolve and affect the structure and/or stability of the fluid, for greater times. This has to emerge from the integration of the full set of Einstein equations but this is out of the scope of the present work. We just wanted to present that the cracking phenomena, which has proven to be related to the problem of structure formation, \cite{MimosoLeDelliouMena2010,MimosoLeDelliouMena2013,LeDelliouEtal2013}, can also be associated to isotropic fluids.    

\section*{Acknowledgment}
 Two of us gratefully acknowledge the support from Vicerrector\'{\i}a de Investigaci\'on y Extensi\'on of the Universidad Industrial de Santander.  AN wants to thank the financial support from the program of J{\'o}venes Investigadores of COLCIENCIAS, Colombia.

\end{document}